# Predicting Financial Crime: Augmenting the Predictive Policing Arsenal

Brian Clifton[1], Sam Lavigne[1], and Francis Tseng[1]

[1] The New Inquiry
https://thenewinquiry.com/

**Abstract.** Financial crime is a rampant but hidden threat. In spite of this, predictive policing systems disproportionately target "street crime" rather than white collar crime. This paper presents the White Collar Crime Early Warning System (WCCEWS), a white collar crime predictive model that uses random forest classifiers to identify high risk zones for incidents of financial crime.

## 1 Introduction

White collar crimes are those "committed by a person of respectability and high social status in the course of his occupation."[2] Examples include the Enron and Bernie Madoff scandals. Incidents of white collar criminality are increasing nationwide, and clearance rates for these crimes are at historic lows[3], despite growing public support for enforcement[4].

Police departments, and predictive policing systems, have historically focused their efforts on reducing "street crimes". However, the development of novel machine learning techniques presents law enforcement with an opportuni-ty to expand their policing efforts into a new domain: high level financial crime.

---

[2] See http://cat.ocw.uci.edu/oo/getPage.php?course=OC0899020&lesson=001&topic=5&page=1

[3] National Public Survey on White Collar Crime, 2010. http://www.nw3c.org/docs/research/2010-national-public-survey-on-white-collar-crime.pdf

[4] Complaints reported to the Federal Trade Commission (FTC) are up 847.22% from 2001 to 2015. https://www.ftc.gov/system/files/documents/reports/consumer-sentinel-network-data-book-january-december-2015/160229csn-2015databook.pdf

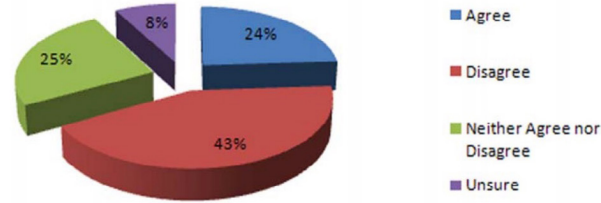

Fig. 1: "Are the Police Devoting Enough Resources to Combat White Collar Crime?" from the 2010 National Public Survey on White Collar Crime.

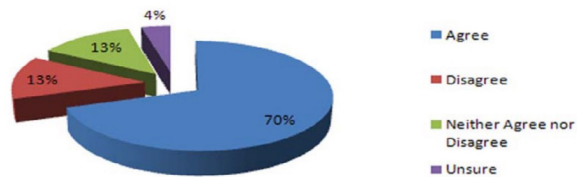

Fig. 2: "White Collar Crime Makes Me Feel Unsafe" from the 2010 National Public Survey on White Collar Crime.

We propose and develop a predictive policing algorithm, the White Collar Crime Early Warning System (WCCEWS), for identifying and assessing the risk of large-scale financial crime at the city block level. Our model achieves an impressive 90.12% accuracy at predicting the activity of white collar crime in a given area.

WCCEWS predicts the likelihood of a white collar crime occurring within a 76m2 square, which is a 197.37% improvement of precision when compared with other predictive policing algorithms[5]. The model is augmented to predict the nature of the white collar crime, as well the severity of the crime (in terms

---

5 See `http://teamcore.usc.edu/projects/security/Muri_publications/Short_JASA_2015.pdf`

of expected fines).

In developing our model, we focused on the physical features of the underlying landscape as described in the risk terrain modeling approach[6] (explained further below), while deprioritizing temporal features. As such, we assert with a high degree of confidence that white collar crimes are occurring continuously in the predicted high-risk zones.

Our model is optimized for growth in the policing of high level financial criminals, with growth measured in higher arrest rates, improved quality of arrests, and higher recovery of funds. The system is designed to be further integrated into tools for citizen policing and awareness, such as the White Collar Crime Risk Zones iOS app, which alerts users when they enter high-risk areas for financial crime.

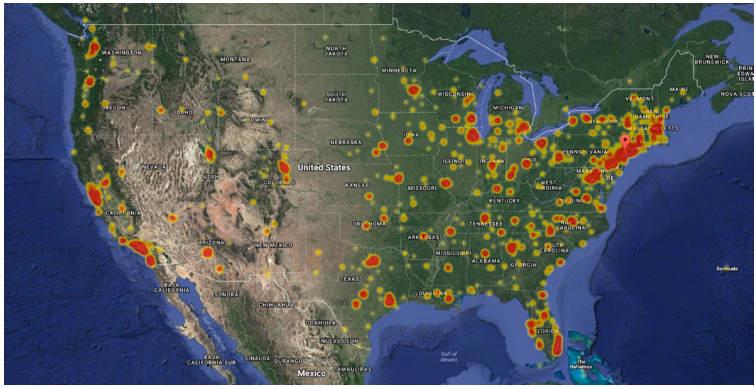

Fig. 3: Financial Crime Risk Surface for the United States.

## 2 Related Work

Our work is inspired by and expands on other predictive policing efforts, such as HunchLab[7], PredPol[8], Hitachi's Predictive Crime Analysis[9], and LexisNexis Risk Solutions' Accurint® Crime Analysis[10], among others. These services over-whelmingly target "street" or "traditional" crime, such as drug-related activity and larceny, overlooking the rich opportunity in targeting financial crimes with their technologies.

---

[6] See http://www.rutgerscps.org/rtm.html

[7] See https://www.hunchlab.com/

[8] See http://www.predpol.com/

[9] See https://www-ssl.intel.com/content/dam/www/public/emea/xe/en/documents/solution-briefs/iot-hitachi-smart-communities-solution-brief.pdf

[10] See http://www.lexisnexis.com/risk/products/government/ac-

These services take the general approach of training classification models with geospatial features and/or historical crime data. They typically focus on predicting crime for a particular geographical region rather than at the individual level[11]. We adopt a similar approach, known as "Risk Terrain Modeling", or RTM, a spatial risk analysis technique first proposed by Les Kennedy and Joel Caplan at Rutgers University[12]. RTM is "used to identify risks that come from features of a landscape and model how they co-locate to create unique behavior settings for crime."[13]

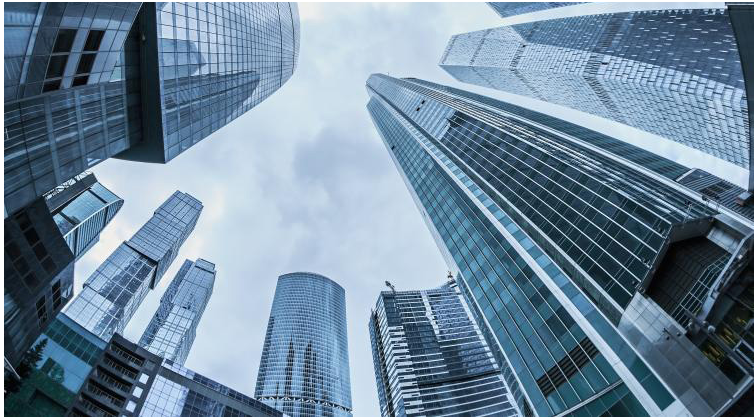

Fig. 4: Example of features in a landscape that create unique behavior settings for white collar criminal activity.

## 3 Data

We collected data provided by the Financial Regulatory Authority (FINRA)[14] to compile incidents of financial malfeasance dating back to 1964. Using these data, we correlated financial crimes to the location of the perpetrating individ-

---

curint-crime-analysis.aspx
[11] See `https://www.gitbook.com/book/teamupturn/predictive-polic-ing/details`
[12] See `http://www.rutgerscps.org/software.html`
[13] See `http://www.rutgerscps.org/rtm.html`
[14] See `https://www.finra.org/`

ual or organization. Financial crimes were geographically clustered according to geohashes computed from these locations.

Numerous other data were incorporated during the feature engineering process. As specified in the risk terrain modeling approach, we compiled geographic data from a variety of sources that contributed to boosting the predictive power of our model. In particular, we looked at: 1) the locations of investment advisers[15]; 2) the geographic distribution of liquor licenses[16]; and 3) the density of tax-exempt organizations[17].

## 3.1 Geohashes

Our geographical data started in the form of street addresses, which we converted to latitude/longitude coordinates with a popular geocoding service[18].

Coordinates are too fine-grained for useful predictions, so we converted our coordinates to *geohashes*. A geohash is a set of characters that all coordinates in some region map to. For example, the coordinates `(40.15, 74)`, `(40.1, 74.1)`, `(40.1, 73.9)` all map to the geohash `txhs` (with a precision level of 4).

The *precision* of a geohash correlates to the size of the region coordinates are mapped to. A more precise geohash represents a smaller region and maps to a longer set of characters (increased precision requires increased specificity, and thus more characters).

For example, when using a precision of 6, those same coordinates map to `txhs7v`, `txhsn5`, `txhs1e`, respectively. Note that the first four characters of each are still txhs.

For our model we use geohashes with a precision level of 7, which map to regions of a $0.076\text{km}^2$ box.

---

[15] Enigma. (2017) Form adv - investment adviser information.
[16] Enigma. (2017) United states liquor licenses.
[17] Enigma. (2017) United states tax-exempt organizations.
[18] See `https://wiki.openstreetmap.org/wiki/Nominatim`

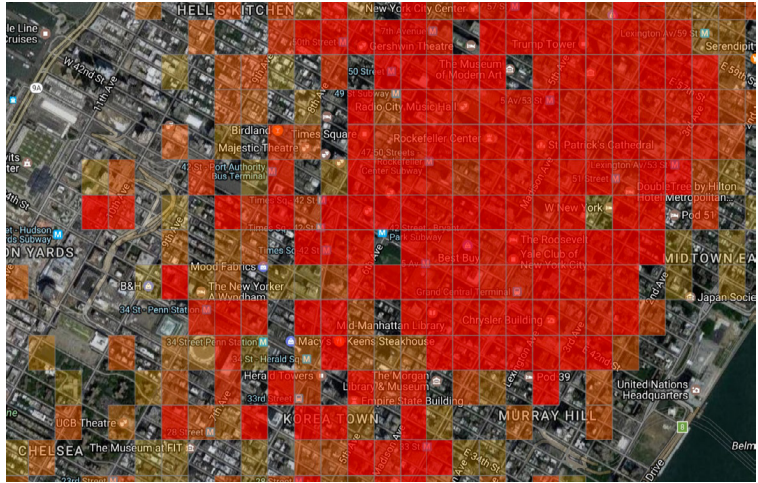

Fig. 5: Geohashes are shown as rectangles. Color intensity shows predicted density of white collar crimes.

## 4 The Model

Our model is composed of three sub-models, each trained to generate a separate prediction:

$M_{crime}$ predicts the probability of any financial crime occurring for a geohash, and is consistent with industry standards for predictive policing applications[19]. We use a forest of decision trees each generated by a bootstrap sample (i.e. a random forest model). The final prediction probability is the average of each tree's predicted probability. This highly interpretable model is similar to the approach HunchLab uses[20]. Our model is trained on the aforementioned data.

$M_{fine}$ predicts the expected fine were a financial crime to take place in a geohash. It is a linear regression model trained on the same auxiliary data, with some additional polynomial features generated from the same data.

$M_{type}$ predicts a distribution over the types of financial crimes likely to occur in

---

[19] See https://www.rand.org/content/dam/rand/pubs/research_reports/RR200/RR233/RAND_RR233.pdf

[20] See https://cdn.azavea.com/pdfs/hunchlab/HunchLab-Under-the-Hood.pdf

a geohash. It is a multi-label (one-vs-rest) random forest model, again trained on the same data.

| Model | Mean Accuracy | Standard Deviation |
|---|---|---|
| $M_{crime}$ | 0.90 | 0.03 |
| $M_{fine}$ | 0.65 | 0.13 |
| $M_{type}$ | 0.46 | 0.80 |

Table 1: Our models' mean accuracies and standard deviations.

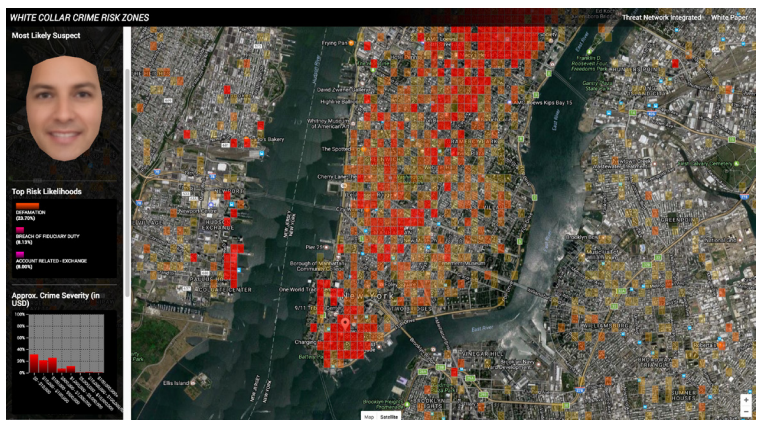

Fig. 6: The WCCEWS user interface. The map shows downtown Manhattan in New York City, NY. Color indicates the predicted density of white collar criminal activity. The left-hand panel shows the Top Risk Likelihoods of the listed crimes occurring within the selected geohash. Below is a histogram indicating predicted Approximate Crime Severity associated with discrete brackets of violation amount in $USD. Finally, the panel lists Potential Offenders operating within the selected geohash, and a generalized white collar criminal subject.

## 5 Conclusion & Future Work

In this paper we have presented our state-of-the-art model for predicting financial crime. By incorporating public data sources with a random forest classifier, we are able to achieve 90.12% predictive accuracy. We are confident that our model matches or exceeds industry standards for predictive policing tools.

Our current model relies solely on geospatial information. It does not consider other factors which may provide additional information about the likelihood of financial criminal activity.

Crucially, our model only provides an estimate of white collar crimes for a particular region. It does not go so far as to identify which individuals within a particular region are likely to commit the financial crime. That is, all entities within high risk zones are treated as uniformly suspicious.

Recently researchers have demonstrated the effectiveness of applying machine learning techniques to facial features to quantify the "criminality" of an individual[21].

We therefore plan to augment our model with facial analysis and psychometrics to identify potential financial crime at the individual level. As a proof of concept, we have downloaded the pictures of 7000 corporate executives whose LinkedIn profiles suggest they work for financial organizations, and then averaged their faces to produce generalized white collar criminal subjects unique to each high risk zone. Future efforts will allow us to predict white collar criminality through real-time facial analysis.

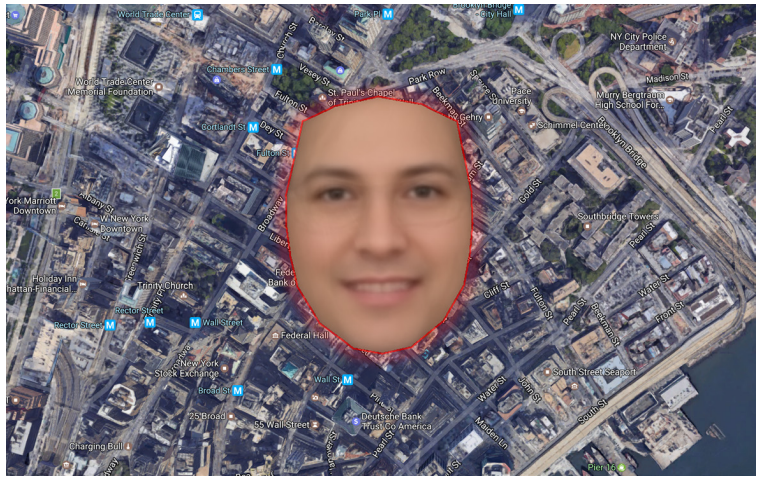

Fig. 7: Predicted White Collar Criminal for `40.7087811, -74.0064149`

---

[21] X. Wu and X. Zhang, "Automated inference on criminality using face images," *CoRR*, vol. abs/1611.04135, 2016.